\documentclass[12pt]{article}
\usepackage[utf8]{inputenc} 
\usepackage{xcolor} 
\usepackage[margin=1.0in]{geometry}
\usepackage{multirow}
\usepackage{array}
\usepackage[normalem]{ulem}
\usepackage{subcaption}
\usepackage{graphicx}
\usepackage{amsmath}
\usepackage{amssymb}
\usepackage[normalem]{ulem}
\usepackage{hyperref}
\usepackage[T1]{fontenc}
\usepackage{charter}
\usepackage{lineno}
\usepackage{longtable}
\usepackage{float}
\usepackage{bigints}
\usepackage{enumitem}
\usepackage{etoolbox}
\usepackage{algorithm}
\usepackage{algpseudocode}

\hypersetup{
  colorlinks   = true, 
  urlcolor     = blue, 
  linkcolor    = blue, 
  citecolor   = blue 
}

\newcolumntype{M}[1]{>{\centering\arraybackslash}m{#1}}

\begin{document}

\begin{center}

{\LARGE\bf{Wave-Assisted Propulsion in Bimodal Sea States: Hydrodynamic Performance and Hydroelastic Tuning}}\\
\vspace{5 mm}

\begin{footnotesize}
Avinash Kumar Pandey, Jung-Hee Seo, Rajat Mittal$^{}$*\\
Department of Mechanical Engineering, Johns Hopkins University, Baltimore, MD 21218, USA.\\
E-mail: mittal@jhu.edu$^{}$*\\
\end{footnotesize}

\end{center}

\begin{abstract}
Wave-assisted propulsion (WAP) systems harvest ocean wave energy to generate propulsive thrust, offering a promising approach for improving endurance and energy efficiency of marine vehicles. Previous studies have focused primarily on monochromatic or unimodal wave conditions, leaving WAP performance in realistic ocean environments largely unexplored. This study investigates the hydrodynamic and hydroelastic response of a submerged flapping hydrofoil operating in bimodal sea states generated by the coexistence of swell and wind-sea wave systems. High-fidelity fluid--structure interaction simulations are performed for representative calm, transitional, and storm conditions, with passive pitching provided through a torsional spring. Simulations show that, despite increased complexity of bimodal wave forcing, propulsion performance follows the same effective peak frequency scaling previously established for monochromatic and unimodal waves, demonstrating the robustness of this scaling framework across a broad range of sea states. The findings further reveal that while the optimal normalized tuning ratio remains within a narrow range, dimensional torsional spring stiffness varies with sea state characteristics, highlighting the need for adaptive hydroelastic tuning to maximize thrust. Overall, the results demonstrate that WAP systems provide a robust means of generating wave-powered thrust under realistic ocean conditions while providing practical guidance for improved design of wave-powered marine propulsion systems.\\


\end{abstract}


\section{Introduction}

\begin{figure}[h!]
\centering
\includegraphics[angle=0, width=\textwidth]{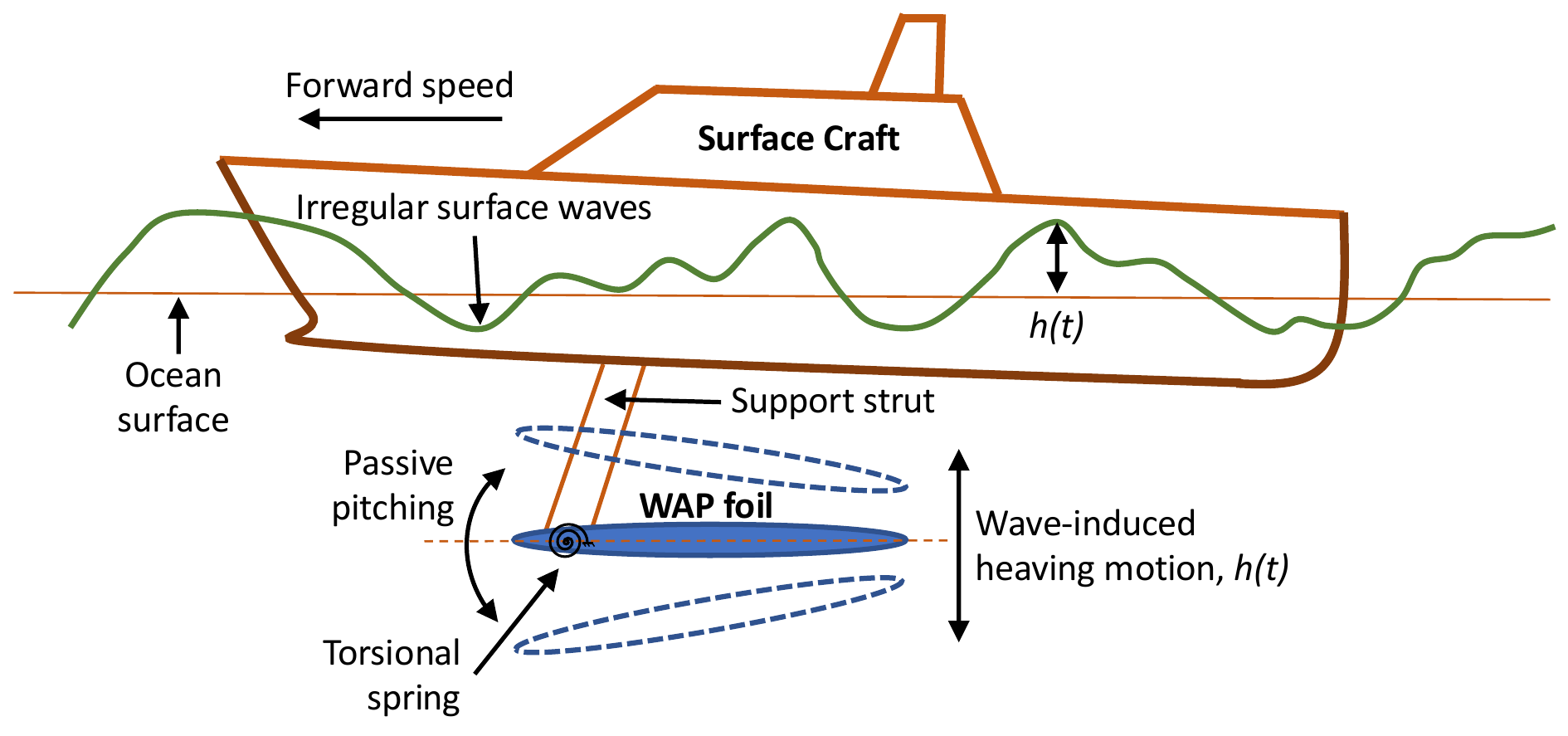}
\caption{Schematic of the wave-assisted propulsion (WAP) system considered in the present study. The system consists of a submerged flapping hydrofoil connected to a surface craft advancing forward. Wave-induced motions of the surface craft generate a prescribed heaving motion of the hydrofoil, while the hydrofoil pitches passively about its hinge under the influence of hydrodynamic loads and a torsional spring.}
\label{wap_schematic}
\end{figure}

Wave-assisted propulsion (WAP) systems have emerged as a promising approach for harvesting energy from ambient ocean waves and converting it into useful propulsive thrust \cite{chan2024wave, feng2024propulsion, feng2025dynamics}. By exploiting the energy naturally available in ocean waves, these systems can supplement conventional propulsion, thereby extending operational range while reducing onboard energy consumption \cite{belibassakis2021numerical, zhang2024dual, liu2016numerical}. Among the various WAP concepts, submerged flapping hydrofoils mounted beneath a surface vessel have attracted particular attention because of their simple mechanical design, structural robustness, and favorable hydrodynamic performance. In the representative configuration shown in Figure~\ref{wap_schematic}, incident waves induce coupled heaving and pitching motions of the hydrofoil, which interacts with the surrounding flow to generate propulsive thrust without requiring direct actuation \cite{wu2020review, raut2024hydrodynamic, raut2025dynamics, kandel2026flapping}.

The hydrodynamics of flapping foils has been studied extensively in the context of biological propulsion \cite{triantafyllou2004review, kumar2026aerodynamic, pai2026propulsive}, energy harvesting \cite{xiao2014review, pandey2023flow, pandey2025dynamics, qi2026inverted}, and marine vehicle design \cite{kandel2026flapping, rozhdestvensky2023recent, wang2024experimental}. Previous studies have shown that the generation of thrust is primarily influenced by the coupling between foil's heaving and pitching motions, the resulting vortex dynamics and the associated formation of leading-edge vortices \cite{raut2025harnessing, eldredge2019leading, pandey2025flow}. Inline with these insights, several recent studies have examined hydroelastic WAP systems in which a submerged foil undergoes wave-induced heaving while pitching passively through elastic or mechanical constraints \cite{raut2024hydrodynamic, wang2024experimental}. Such systems have been shown to produce significant thrust over a range of operating conditions and sea states.


A key aspect of modeling WAP systems is the representation of ocean waves that drive the hydrofoil motion. Early studies typically simplified the wave environment by considering regular monochromatic waves, resulting in either sinusoidal \cite{triantafyllou1993optimal, anderson1998oscillating, wu2020review, raut2025dynamics} or non-sinusoidal \cite{qi2019effects} hydrofoil kinematics. Although such idealized wave models have provided valuable insights into thrust generation mechanisms and the associated flow physics, they do not adequately represent realistic ocean conditions. To address this limitation, several studies have incorporated irregular wave forcing through spectral representations of ocean waves. In this context, time domain simulation frameworks for wave driven systems have been widely developed in the wave energy literature, where irregular wave fields are represented through spectral decomposition into harmonic components and directly integrated in time domain hydrodynamic models \cite{holthuijsen2010waves, falcao2010hydrodynamic}. For example, Romanowski et al. \cite{romanowski2019development} and L{\"u}nser et al. \cite{lunser2022influence} employed JONSWAP wave spectrum to model unimodal irregular sea states. Other approaches, such as the WaveMIMO methodology \cite{oleinik2025numerical, machado2021wavemimo}, reconstruct statistically equivalent time series realizations from prescribed wave spectra. Methods for generating multidirectional irregular waves have also been developed \cite{ha2013generation, wang2019multi}. More recently, irregular wave forcing has been applied to WAP systems, with Raut et al. \cite{raut2025hydrodynamics} demonstrating that spectral bandwidth and intermittent high-energy events can significantly influence foil kinematics and propulsion performance. However, these investigations have primarily considered unimodal wave spectra characterized by a single dominant frequency, which is representative of a swell-dominated sea state. Consequently, the performance of these systems in more complex sea states, such as those generated by strong wind-wave interaction, which are frequently encountered in realistic ocean environments, remains largely unexplored.

\begin{figure}[h!]
\centering
\includegraphics[angle=0, width=\textwidth]{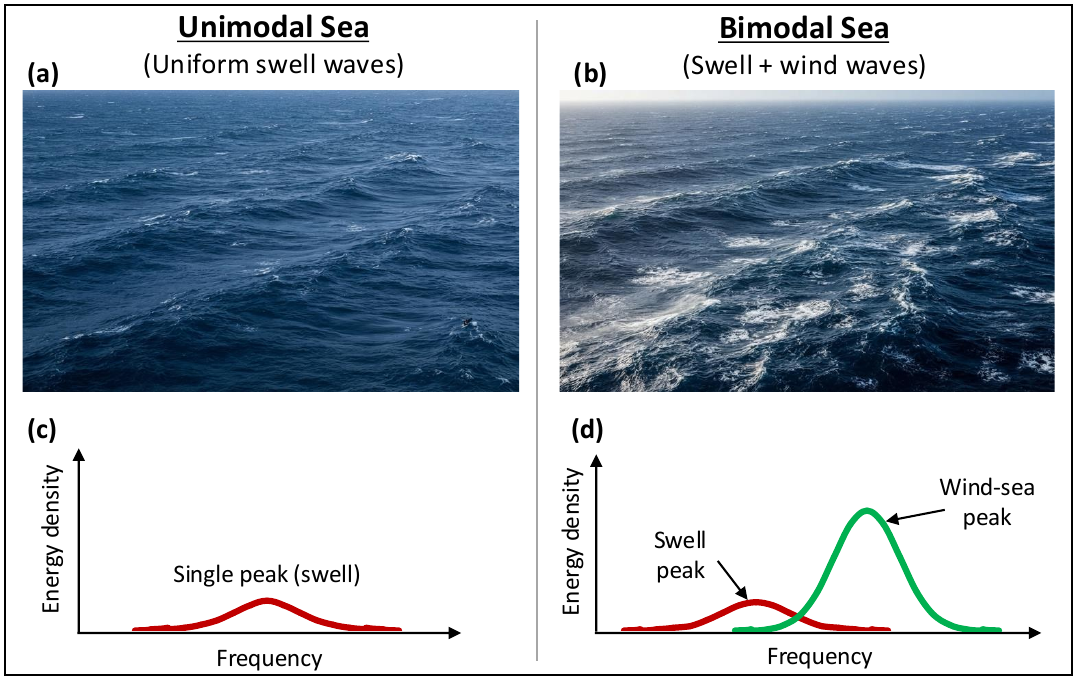}
\caption{Illustration of unimodal and bimodal sea states: (a, b) Visualization of the sea surface in both sea states \cite{google2026gemini}, and (c, d) representative wave energy spectra where the unimodal sea state is characterized by a single dominant spectral peak and a single characteristic wave timescale, whereas the bimodal sea state contains distinct swell and wind-sea components that introduce multiple energetic timescales.}
\label{schematic_bimodal}
\end{figure}

While most studies have focused on unimodal wave spectra representative of swell-dominated seas, practical wave-assisted propulsion (WAP) systems must operate reliably across the broad range of sea states encountered in the open ocean. Measurements show that natural ocean environments frequently exhibit multiple energetic peaks arising from the coexistence of locally generated wind waves (higher frequency, shorter wavelength) and remotely generated swell waves (lower frequency, longer wavelength) \cite{rashmi2013co, chen2015practical, vettor2020global, wang2024effect}, as illustrated in Figure \ref{schematic_bimodal}(a). These bimodal sea states are common in mixed seas, transitional weather, and post-storm conditions. The interaction of the wind-wave and swell components produces wave fields with multiple characteristic frequencies and substantially more complex temporal variations in surface elevation, velocity, and acceleration than unimodal seas (Figure \ref{schematic_bimodal}(b)). Understanding the response of WAP systems to such conditions is therefore essential for assessing their robustness and ensuring reliable propulsion across realistic operating environments. Despite their prevalence and practical importance, the implications of bimodal sea states for wave-assisted propulsion remain largely unexplored.

This inherent complexity of bimodal sea states introduces challenges in establishing reliable scaling relationships for WAP systems. In a unimodal spectrum, quantities such as the spectral peak frequency can often be used to characterize the dominant forcing timescale experienced by the foil \cite{raut2025hydrodynamics}. In bimodal seas, however, no unique dominant frequency exists \cite{portilla2019statistical, vettor2020global, hegermiller2017multimodal} (see Figure \ref{schematic_bimodal}(c)).  Consequently, it is unclear whether conventional frequency-based scaling parameters remain valid for predicting propulsion performance or for selecting optimal parameters for the WAP foil such as spring stiffness and pitch limiting characteristics. Moreover, interactions between swell and wind-sea components may produce intermittent high-velocity heaving events \cite{rashmi2013co, fan2014simulated}, which can substantially alter foil kinematics, vortex formation, and thrust production.

The importance of such intermittent high-velocity motions is further supported by recent findings under unimodal irregular wave forcing. Raut et al. \cite{raut2025hydrodynamics} demonstrated that an elliptic WAP foil subjected to a unimodal irregular wave spectrum can outperform an equivalent sinusoidal excitation with the same total spectral energy. This enhancement was attributed to intermittent high-velocity events that increase the irregular wave-based Strouhal number, $\text{St}_w$ (defined using the ensemble-averaged RMS heave velocity to characterize the effective wave forcing, see Equation (\ref{stw_eqn})), thereby promoting more effective thrust generation. These findings suggest that WAP performance is governed not only by the total wave energy available but also by its temporal and spectral distribution, which determines the resulting heave–pitch kinematics. Whether comparable or amplified effects occur in bimodal sea states, where the interaction between swell and wind-sea components introduces additional characteristic timescales and velocity fluctuations, remains an open question.

To address this gap, the present study investigates the performance of WAP systems in bimodal sea states. The coexistence of swell and wind-sea components introduces fundamental questions regarding hydrofoil kinematics, the applicability of conventional frequency-based performance metrics, and the development of robust hydroelastic tuning strategies. Specifically, this study addresses the following questions:

\begin{itemize}[noitemsep, left=0pt, labelindent=0pt]
\item How does the distribution of wave energy between swell and wind-sea components affect foil kinematics and propulsive thrust?
\item Can propulsion performance in bimodal seas be characterized using a single effective peak frequency despite the presence of two dominant wave components?
\item How does hydroelastic tuning influence WAP performance, and how does the optimal tuning vary across different bimodal sea states?
\item Does spectral bimodality give rise to distinct hydrodynamic response regimes, and what are their implications for WAP system design?
\end{itemize}

These questions are investigated using high-fidelity fluid--structure interaction simulations of a submerged hydrofoil subjected to heaving motions generated from synthetic bimodal wave spectra (Figure~\ref{wap_schematic}). The foil is free to pitch through a torsional spring, enabling passive hydroelastic tuning via a spring--limiter mechanism in which the effective torsional stiffness is systematically varied. Representative sea states spanning calm, transitional, and storm conditions are considered by systematically varying the relative energy content of the swell and wind-sea components (Figure~\ref{schematic_bimodal}(c)). Propulsion performance is then analyzed in terms of foil kinematics, thrust generation, hydroelastic response, and the underlying vortex dynamics.


\section{Methodology}

\subsection{Wave-assisted propulsion (WAP) system}
The WAP system considered in the present study is shown schematically in Figure \ref{wap_schematic}. The configuration consists of a submerged hydrofoil which is modeled as being connected to a notional surface craft through a linkage that transfers wave-induced heaving motion to the foil. Note that the surface craft is not included in the model. The heave displacement is prescribed based on calculated bimodal sea surface elevations. In addition, the foil is allowed to undergo passive pitching (i.e., flow-induced rotation) about its pitching axis through an attached torsional spring, which acts to limit the overall pitching motion \cite{raut2024hydrodynamic}, with the resulting pitching response governed by the balance between hydrodynamic moments and elastic restoring moments. By varying the spring stiffness, the hydroelastic response of the foil can be tuned. Thus, the foil motion consists of a prescribed heaving component coupled with a passively generated pitching response.

\subsection{Modeling bimodal ocean wave motion} \label{bimodal_model_s}

The heaving motion prescribed to the hydrofoil, denoted as $h(t)$ in Figure \ref{wap_schematic}, is derived from modeling of irregular ocean waves. The wave field is modeled using a parametric spectral approach, where the spectral energy density $S(f)$ is defined as a function of frequency, $f$. Previous work \cite{raut2025hydrodynamics} employed a unimodal Bretschneider spectrum \cite{bretschneider1959wave} to represent irregular sea states with a single dominant spectral peak. In the present study, this approach is extended to bimodal sea states \cite{ewans2006estimation, mackay2016unified} by superimposing distinct wind-sea and swell spectral components. Such mixed sea conditions are frequently observed in realistic ocean environments \cite{wang2001operational}.

Each component of the bimodal spectrum, namely wind-sea and swell contributions, is represented using the Joint North Sea Wave Project (JONSWAP) spectral formulation \cite{hasselmann1973measurements}. The total wave spectrum is then constructed by superimposing the individual spectral components as follows,
\begin{equation}
S(f) = S_o \, g^2 (2\pi)^{-4} f^{-5}
\exp\left[-\frac{5}{4}\left(\frac{f_p}{f}\right)^4\right]
\gamma^{\exp\left[-\frac{(f-f_p)^2}{2\sigma^2 f_p^2}\right]} ,
\end{equation}
where $S(f)=S^*(f^*) {U^*_\infty}/{C^{*3}}$ is the wave energy spectral density, $g$ denotes the gravitational acceleration, $f_p$ (=$f_p^*C^*/U_\infty^*$) is the peak frequency, $\gamma$ is the peak enhancement factor, and $\sigma$ takes values $\sigma=0.07$ for $f \le f_p$ and $\sigma=0.09$ for $f > f_p$. The parameter $S_o$ is determined such that the integrated spectral energy corresponds to a prescribed significant wave height $H_s$ (=$H_s^*/C^*$).

The non-dimensional spectral energy, $E$ is related to the zeroth spectral moment through,
\begin{equation}
E = \int_{0}^{\infty} S(f) df
\end{equation}
which is connected to the significant wave height via,
\begin{equation}
H_s = 4\sqrt{E}.
\end{equation}

The composite wave spectrum is obtained by superimposing the wind-sea ($S_w$) and swell ($S_s$) components  \cite{boukhanovsky2009modelling, akbari2019double, rossi2021investigation, liang2022modeling} as follows,
\begin{equation}
S_{\text{total}}(f) = S_w(f) + S_s(f).
\label{spec_dens_tot}
\end{equation}
Consequently, the resulting wave field contains two energetic spectral peaks corresponding to the swell and wind-sea waves. These components are characterized by distinct peak frequencies ($f_{p,S}$ and $f_{p,W}$), total spectral energies ($E_S$ and $E_W$), and significant wave heights ($H_{S,S}$ and $H_{S,W}$). The resulting \emph{double} JONSWAP wave spectrum introduces multiple forcing timescales and wave energy pathways that are absent in unimodal sea states, thereby creating a more realistic representation of ocean wave excitation \cite{wang2001operational} for assessing WAP systems.

\subsubsection{Effective peak spectral frequency} \label{eff_freq_s}

For unimodal spectra, the spectral peak frequency provides a convenient measure of the dominant wave timescale. In bimodal spectra, however, the presence of two energetic peaks associated with the swell and wind-sea components precludes the existence of a unique characteristic frequency. To facilitate comparisons across different bimodal sea states, an effective peak spectral frequency, $f_{\mathrm{eff}}$, can be defined as an energy weighted average of the swell and wind-sea peak frequencies as follows,
\begin{equation}
f_{\mathrm{eff}} = \frac{E_S f_{p,S}+E_W f_{p,W}} {E_S+E_W},
\label{eff_freq_eq}
\end{equation}
where $E_S$ and $E_W$ denote the total spectral energies associated with the swell and wind-sea components, respectively. This definition accounts for both spectral peaks while weighting their contributions according to their relative energy content. Consequently, $f_{\mathrm{eff}}$ provides a single characteristic timescale for the bimodal wave field and serves as a convenient parameter for characterizing the wave forcing experienced by the hydrofoil.

\subsubsection{Characterization of sea states} \label{ss_char_s}
To systematically investigate the influence of bimodal wave forcing on the hydrodynamic response of the WAP system, sea states are categorized according to the relative energy content of the swell and wind-sea components comprising the composite wave spectrum. The total spectral energy of a bimodal sea state is given by,
\begin{equation}
E_\text{total} = \int_0^\infty S_\text{total}(f) df,
\end{equation}
where $S_\text{total}(f)$ denotes the composite spectral density defined in Equation (\ref{spec_dens_tot}). The total energy may be decomposed into swell and wind-sea contributions as follows,
\begin{equation}
E_\text{total} = E_S + E_W,
\end{equation}
where $E_S$ and $E_W$ are the total spectral energies associated with the swell and wind-sea components, respectively. To quantify the relative dominance of the two wave systems, a swell to wind-sea energy ratio is defined as follows,
\begin{equation}
\lambda_E = \frac{E_S}{E_W} .
\end{equation}
As the total wave energy scales with the square of significant wave height \cite{holthuijsen2010waves}, $\lambda_E$ may be obtained directly by,
\begin{equation}
\lambda_E = \frac{H_{s,S}^2}{H_{s,W}^2} .
\label{energy_ratio_E}
\end{equation}

Notably, larger values of $\lambda_E$ correspond to swell-dominated conditions, whereas smaller values indicate the dominance of the wind-sea component. Based on this parameter, the following sea states are considered in the present work,
\begin{itemize}
\item \textbf{Sea State 1 (Calm Sea):} $\lambda_E > 1$ (Swell dominated).
\item \textbf{Sea State 2 (Transitional Sea):} $\lambda_E = 1$ (Equal swell and wind-sea spectral energy).
\item \textbf{Sea State 3 (Storm Sea):} $\lambda_E < 1$ (Wind-sea dominated).
\end{itemize}

\begin{figure}[h!]
\centering
\includegraphics[angle=0, width=\textwidth]{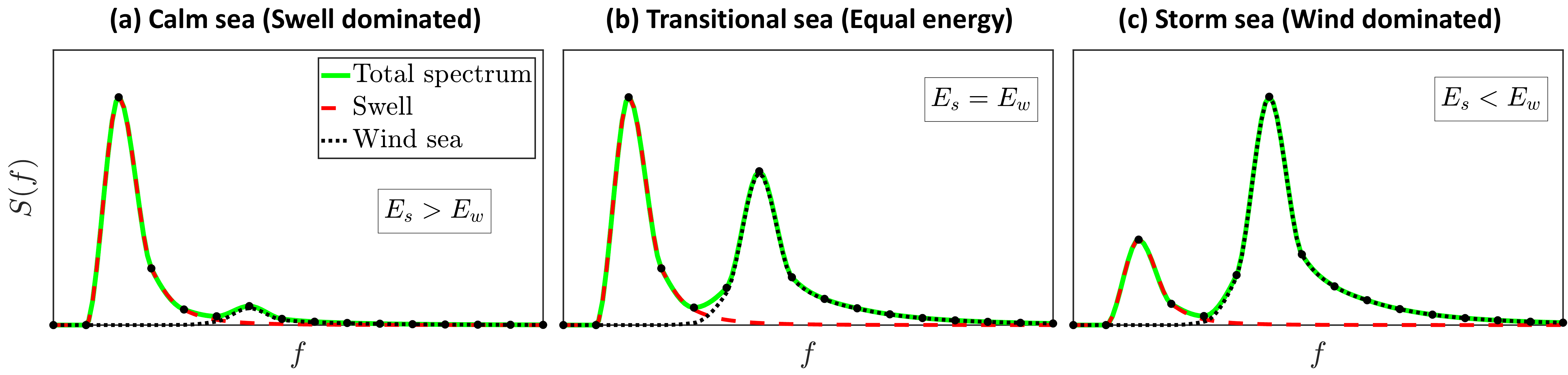}
\caption{Representation of the three bimodal sea states considered in this study, defined by varying swell to wind-sea energy ratios $\lambda_E$ : (a) Calm sea state (swell-dominated, $\lambda_E > 1$), (b) transitional sea state with equal swell and wind-sea energy contributions ($\lambda_E = 1$), and (c) storm sea state (wind-sea dominated, $\lambda_E < 1$). The corresponding spectral energy distributions illustrate the relative contribution of the two JONSWAP components (swell and wind-sea) forming the composite bimodal wave spectrum.}
\label{ss_characterization}
\end{figure}

Representations of the corresponding spectral energy distributions are shown in Figure \ref{ss_characterization}. By varying the total spectral energy between the two spectral peaks while maintaining a bimodal wave structure, these sea states provide a systematic framework for examining the influence of realistic sea conditions on hydrofoil kinematics and propulsive performance.

\subsection{Hydrofoil kinematics}

\subsubsection{Prescribed heaving motion based on wave surface elevation} \label{pres_heave}

The spectral quantities introduced in Section \ref{bimodal_model_s} provide a frequency domain description of the bimodal sea state and its associated energy distribution. However, the hydrodynamic response of the WAP system is governed by the time-varying motion experienced by the hydrofoil. Accordingly, the composite wave spectrum is transformed into a time domain elevation of the sea surface, which is subsequently used to prescribe the hydrofoil heaving motion.

For numerical implementation, the frequency domain is discretized into $N$ equally spaced frequency bins with spacing $\Delta f$. The sea surface elevation is then reconstructed using linear spectral superposition, whereby the irregular wave field is represented as the sum of harmonic components. The resulting wave surface elevation ($h=h^*/C^*$) is expressed as,
\begin{equation}
h(t) = \sum_{j=1}^{N} a_j
\cos \left( 2\pi f_j t + \phi_j \right),
\label{sur_ele}
\end{equation}
where $f_j$ denotes the discretized frequencies and $\phi_j$ are randomly distributed phases in the interval $[0,2\pi]$. The amplitude of each harmonic component is determined from the spectral density as,
\begin{equation}
a_j = \sqrt{2 S(f_j) \Delta f}.
\end{equation}
The reconstructed wave surface elevation, $h(t)$ in Equation (\ref{sur_ele}), is subsequently used to prescribe the hydrofoil heaving motion. Specifically, the hydrofoil is assumed to follow the wave-induced vertical displacement of the surface craft, such that its heave kinematics is directly determined by the reconstructed wave surface elevation signal in Equation (\ref{sur_ele}).

\subsubsection{Flow-induced pitching with spring limiter} \label{passive_pitch}

\begin{figure}[h!]
\centering
\includegraphics[angle=0, width=0.9\textwidth]{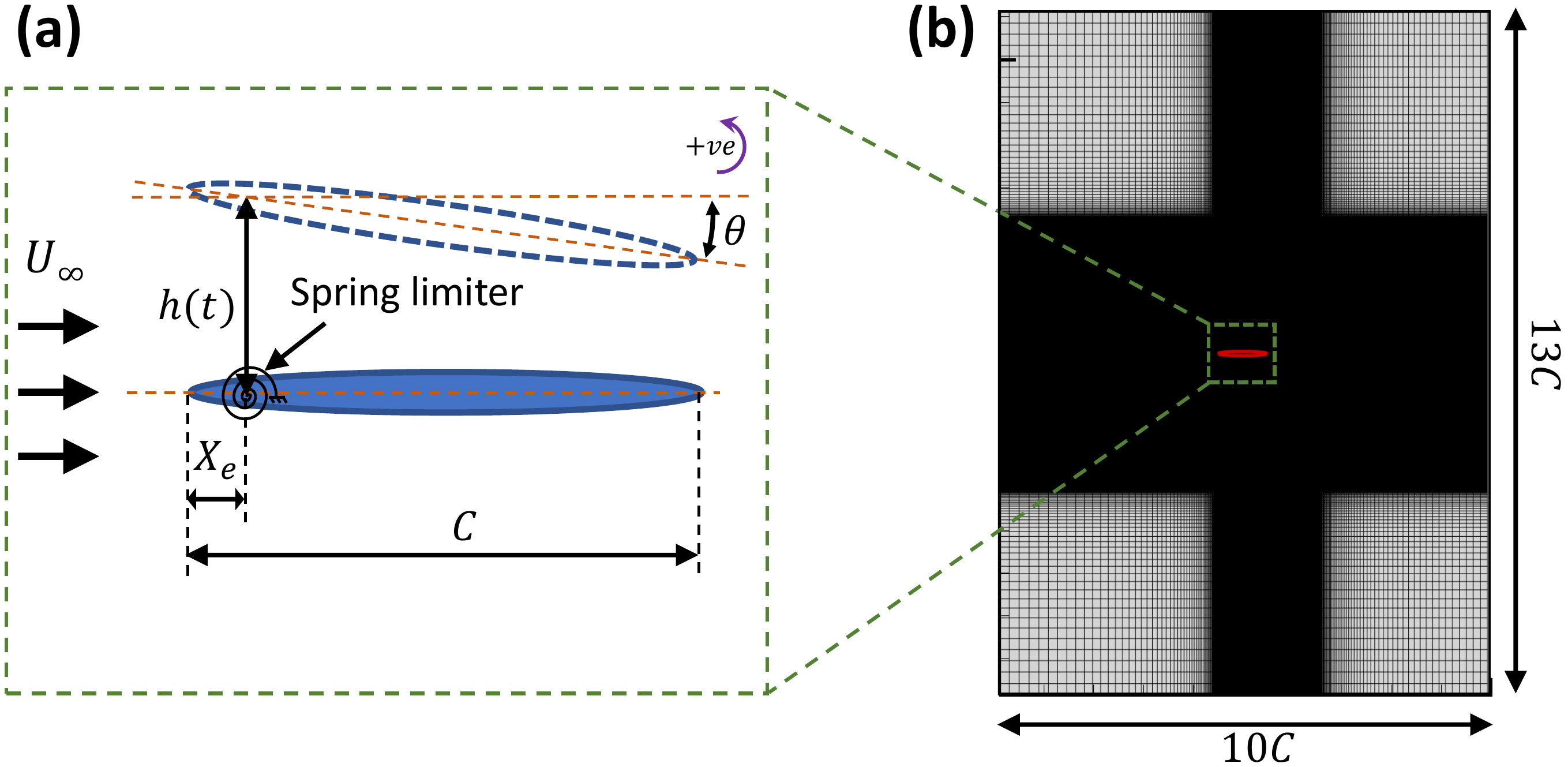}
\caption{(a) Schematic of the wave-assisted propulsion hydrofoil undergoing prescribed heaving motion, $h(t)$, and passive pitching motion, $\theta$, regulated by a spring limiter located at the pitch axis $X_e$. (b) Computational domain and Cartesian grid employed for the immersed boundary simulations.}
\label{problem_grid}
\end{figure}

The passive pitching response of the hydrofoil is regulated with a torsional spring mounted at the pitch axis (Figure \ref{problem_grid}(a)). The spring acts as a pitch constraining mechanism, herein referred to as the \textit{spring limiter}, and introduces compliance into the hydrofoil dynamics through the balance between hydrodynamic moments and elastic restoring torque. Such compliant pitch regulation has previously been shown to enhance thrust generation in wave-assisted propulsion systems by enabling favorable phase relationships between the heaving and pitching motions \cite{raut2025hydrodynamics}.

The passive pitching motion of the hydrofoil is governed by the equation of a forced angular spring--mass oscillator as follows,
\begin{equation}
I \ddot{\theta} + K \theta = C_M,
\label{eqn_pitch}
\end{equation}
where $\theta$ is the instantaneous pitch angle. Moreover, $I$ $(=2I^{*}_k/\rho^*_f C^{*4} b^*)$, $K$ ($= 2K^* / \rho_f^* U^{*2}_\infty C^{*2} b^*$), and $C_{M}$ $ (={2 M_{\theta}^{*}} / {\rho^*_f U^{*2}_\infty C^{*2} b^*})$ are the non-dimensional moment of inertia, spring stiffness and hydrodynamic pitching moment coefficient, respectively. Here, $M_{\theta}^{*}$, $K^*$ and $b^*$ are the dimensional hydrodynamic pitching moment, spring stiffness and spanwise width, respectively. $C_{M}$ is calculated about the pitch axis located at $X_e$ distance from the leading edge of hydrofoil (Figure \ref{problem_grid}(a)).

In previous studies involving regular waves \cite{raut2024hydrodynamic} or unimodal irregular waves \cite{raut2025hydrodynamics}, the spring stiffness was conveniently characterized through the frequency ratio $f_\theta/f_p$, where $f_p$ denotes the dominant spectral peak frequency \cite{raut2025hydrodynamics}. However, for bimodal sea states in the present work, the presence of distinct swell and wind-sea peaks precludes the existence of a unique characteristic frequency. Consequently, the spring limiter is parameterized using the effective peak spectral frequency introduced in Section \ref{eff_freq_s}. The pitch governing Equation (\ref{eqn_pitch}) may therefore be expressed as,

\begin{equation}
\ddot{\theta} + \left( 2\pi \frac{f_\theta}{f_{\mathrm{eff}}} \text{St}_{\mathrm{eff}} \right)^2 \theta = \frac{C_M}{I},
\label{eqn_pitch_st}
\end{equation}
where $\text{St}_{\mathrm{eff}}$ ($ = {f_{\mathrm{eff}} C^*}/{U^*_\infty}$) is the effective peak frequency-based Strouhal number and $f_\theta = 1/2\pi \sqrt{K/I}$ is the natural frequency of the spring mass system. Moreover, ${f_\theta}/{f_{\mathrm{eff}}}$ is the non-dimensional frequency ratio, which therefore serves as the primary parameter governing the spring limiter response. Physically, this ratio represents the degree of tuning between the natural dynamics of the compliant hydrofoil and the characteristic timescale of the bimodal wave excitation. Small values correspond to relatively compliant spring systems, whereas larger values represent increasingly stiff pitch constraints.

Notably, defining an appropriate characteristic velocity and frequency for evaluating the Strouhal number under irregular wave forcing requires additional consideration. Unlike regular waves, where the characteristic heave velocity is uniquely determined by the prescribed wave amplitude and frequency, irregular waves produce stochastic heave motions with continuously varying amplitudes and frequencies, precluding a unique instantaneous velocity scale. Since the hydrodynamic response is governed primarily by the heave velocity experienced by the flapping foil, the root-mean-square (RMS) heave velocity is adopted as a representative measure of the excitation. Following Raut et al. \cite{raut2025hydrodynamics}, the corresponding effective wave-based Strouhal number is therefore defined as,


\begin{equation}
St_w = \frac{\sqrt{2} \dot{h}_{\mathrm{rms}}}{\pi U_\infty},
\label{stw_eqn}
\end{equation}
where $\dot{h}_{\mathrm{rms}}$ is the ensemble-averaged RMS heave velocity of the prescribed bimodal wave-induced motion. This definition provides a physically meaningful measure of the effective wave forcing intensity and is used throughout the present study to characterize the operating conditions under irregular wave excitation.

\subsection{Fluid-structure interaction and numerical solver}

\subsubsection{Fluid flow}
The fluid flow is modeled as an incompressible viscous flow governed by the Navier-Stokes equations, employed in their non-dimensional form as follows,
\begin{equation}
\nabla \cdot \mathbf{u}=0,
\label{cont}
\end{equation}
\begin{equation}
\frac{\partial \mathbf{u}}{\partial t} + \mathbf{u}\cdot\nabla\mathbf{u} = -\nabla p + \frac{1}{Re} \nabla^2 \mathbf{u}.
\label{NS}
\end{equation}
where $u$ and $p$ are the fluid velocity and pressure, respectively, and $Re$ ($=\rho_f^* U_\infty^* C^*/\mu^*$) represents the Reynolds number. For the two-dimensional computational domain considered herein, $i,j=1,2$.

The flow simulations are performed using a sharp interface, immersed boundary solver \texttt{ViCar3D} \cite{mittal2008versatile, seo2011sharp, kumar2026aerodynamic, pandey2025flow}. The solver employs a non body-fitted Cartesian grid while maintaining a sharp representation of the immersed boundary. For flow field solution, Equation (\ref{NS}) is marched in time using the fractional step method comprising three sub-steps: (i) computation of an intermediate velocity field by solving the modified momentum equations, (ii) pressure correction through the solution of a pressure Poisson equation to enforce the incompressibility constraint given by Equation (\ref{cont}), and (iii) update of the velocity and pressure fields to the next time level.


\subsubsection{Coupling}

A partitioned fluid--structure interaction (FSI) framework is employed to couple the fluid and structural dynamics. At each time step, $t=t^n$, the current hydrofoil configuration is used to impose velocity boundary conditions at the fluid--structure interface through the ghost cell methodology \cite{mittal2008versatile}. The incompressible flow Equations (\ref{cont}) and (\ref{NS}) are then solved to obtain the updated pressure and velocity field, from which the hydrodynamic forces and moments acting on the immersed body are evaluated. The resulting hydrodynamic moment coefficient, $C_M$, is subsequently introduced into Equation (\ref{eqn_pitch_st}) to determine the hydrofoil pitch angle, $\theta^{n+1}$, at the next time step. The heave displacement of the hydrofoil is prescribed independently from the reconstructed wave surface elevation signal, $h(t)$, using Equation (\ref{sur_ele}). The updated pitch angle and heave position are then used to update hydrofoil position for $t=t^{n+1}$, and the procedure is repeated throughout the simulation. An explicit coupling strategy is adopted between the fluid and structural solvers, which has been demonstrated to remain stable for systems with sufficiently large structural inertia \cite{menon2019flow}.

The \texttt{ViCar3D} solver employed in the present study has been extensively verified previously for a broad range of FSI problems, including stationary and moving boundary configurations \cite{menon2019flow, raut2024hydrodynamic, raut2025hydrodynamics, pandey2023flow, pandey2025flow}, as well as flexible zero thickness structures such as plates and membranes \cite{mittal2008versatile, zhou2024effect, kumar2025computational, prakhar2025bioinspired, zhou2025hydrodynamically}.

\subsection{Simulation setup} \label{sim_setup}

The numerical configuration employed in the present study is illustrated in Figure \ref{problem_grid}. A two-dimensional computational domain of dimensions $10C \times 13C$ is adopted for all simulations. An elliptic hydrofoil of aspect ratio 25:2 is positioned such that its leading edge is located at a distance of $4.5C$ from the inlet boundary. Within the immersed boundary framework, the Cartesian mesh remains fixed throughout the simulation \cite{mittal2023origin}, allowing the hydrofoil to undergo prescribed kinematics without the need for grid deformation or remeshing. A locally refined uniform mesh of size $\Delta x = \Delta y = 0.0045C$ is employed around the hydrofoil to accurately resolve boundary layer development, vortex shedding, and hydrodynamic loading, while a progressively coarser grid is used away from the immersed body to improve computational efficiency. The domain dimensions and mesh resolution are selected based on previously established verification and grid/domain independence studies for the same WAP configuration under similar operating conditions \cite{raut2025hydrodynamics}.

For boundary conditions, a uniform velocity $U_{\infty}$ is imposed at the inlet. Free stream velocity boundary conditions are specified along the upper and lower boundaries, whereas a fully developed flow condition is applied at the outlet. Neumann boundary conditions are prescribed for pressure correction on all boundaries.

At the start of each simulation, the hydrofoil is initialized at its mean heave position with zero pitch angle, while both translational and angular velocities are set to zero. At later time steps, the hydrofoil undergoes a prescribed heaving motion obtained from the reconstructed bimodal wave signal described in Section \ref{pres_heave}, while its pitching response is determined by the FSI framework through Equation (\ref{eqn_pitch_st}).

Three independent phase realizations ($R_1$, $R_2$ and $R_3$) are generated for each simulation data point to ensure statistical convergence of the propulsion metrics. Each simulation is performed for a sufficiently long duration such that initial transients decay and statistically stationary behavior is attained. The hydrodynamic and propulsion metrics reported in the subsequent sections are obtained by averaging over both time and the ensemble of independent wave realizations.

\subsection{Parametric space} \label{param_space_s}


The numerical framework described above is characterized by a combination of hydrodynamic, structural, and wave spectral parameters. All simulations in this study are performed at a Reynolds number of $Re=10^4$, with a hydrofoil inertia of $I=0.28$ and a pitch axis location of $X_e=0.1$, consistent with Raut et al. \cite{raut2025hydrodynamics}. The bimodal wave spectral parameters are adopted from the representative swell and wind-sea conditions reported by Piscopo et al. \cite{piscopo2025assessment}, ensuring that the prescribed wave forcing is representative of realistic ocean sea states. Accordingly, the peak frequencies and peak enhancement factors are fixed at $f_{p,S}=0.05$, $f_{p,W}=0.16$, and $\gamma_S=1.2$, and $\gamma_W=3.3$, respectively.

To discretize the continuous wave spectrum, a uniform frequency bin width of $\Delta f = f_{p,W}/6$ is employed. Because the discrete frequency components are integer multiples of $\Delta f$, the reconstructed irregular wave signal is periodic over the fundamental reconstruction period of $T = 1/\Delta f = 6/f_{p,W} = 37.5$. Accordingly, all cycle-averaged quantities are evaluated over this reconstruction period for all cases in this paper. Accordingly, all cycle-averaged quantities reported in this work are computed over this fixed period $T=37.5$.

The primary objective of this work is to investigate the influence of bimodal wave forcing under different sea states on the propulsion characteristics of the WAP system in Figure \ref{problem_grid}. To isolate the effect of wave energy while preserving the underlying spectral characteristics, the peak frequencies of the swell and wind-sea components are kept constant and only their significant wave heights are varied. This choice is motivated by the fact that wave energy is proportional to the square of the significant wave height \cite{holthuijsen2010waves}. Therefore, varying the significant wave heights systematically modifies the energy content without changing the characteristic frequencies.

Accordingly, the parameters varied in the present study are the relative swell to wind-sea energy ratio, $\lambda_E$, and the spring limiter frequency ratio, $f_\theta/f_{\mathrm{eff}}$. The frequency ratio is varied over the range $2 \leq f_\theta/f_{\mathrm{eff}} \leq 10$ to examine the influence of hydroelastic tuning under different bimodal sea states characterized in Section \ref{ss_char_s}. Three representative classes of bimodal sea states are considered: a swell-dominated calm sea (SS1), a transitional sea with comparable swell and wind-sea energy contributions (SS2), and a wind-dominated storm sea (SS3). Within the calm sea category, three additional sub-cases are examined by progressively increasing the wind-sea significant wave height while maintaining identical swell characteristics. Following our previous work \cite{raut2025hydrodynamics}, for each sea state and spring limiter configuration, three independent random phase realizations of the bimodal spectrum ($R_1$, $R_2$ and $R_3$) are generated to account for the stochastic nature of irregular ocean waves. The complete set of wave conditions considered in the present study is summarized in Table \ref{params_table}. Notably, $\text{St}_w$ increases with sea states (see Table \ref{params_table}) due to increasing contribution of the higher frequency wind-sea component ($f_{p,W} > f_{p,S} $), which increases $\dot{h}_{\mathrm{rms}}$, and consequently $\text{St}_w$, as given by Equation (\ref{stw_eqn}).
\begin{table}[h!]
\centering
\caption{Bimodal sea states considered in the present study. For each sea state, the spring limiter frequency ratio spans $2 \leq f_\theta/f_{\mathrm{eff}} \leq 10$, with three independent random phase realizations simulated for each sea state and frequency ratio combination.}
\label{params_table}
\setlength{\tabcolsep}{12pt} 
\begin{tabular}{ccccccc}
\hline
Case &
$H_{s,S}$ &
$H_{s,W}$ &
$E_\text{total}$ &
$\lambda_E$ &
$\text{St}_w$ &
Sea state \\
\hline
SS1a & 1.5 & 0.5 & 0.156 & 9.0 & 0.103 & Calm \\
SS1b & 1.5 & 0.672 & 0.169 & 4.98 & 0.12 & Calm \\
SS1c & 1.5 & 0.993 & 0.202 & 2.28 & 0.16 & Calm \\
SS2  & 1.5 & 1.5 & 0.281 & 1.0 & 0.228 & Transitional \\
SS3  & 1.5 & 3.0 & 0.703 & 0.25 & 0.437 & Storm \\
\hline
\end{tabular}
\end{table}

\subsection{Data analysis}

\subsubsection{Performance metric}

The hydrodynamic performance of the WAP system is quantified by the coefficient of thrust, $C_T$, defined as follows,
\begin{equation}
    C_T = \cfrac{2F_T^*}{\rho^*_f U_\infty^{*2} C^* b^*},
    \label{coeff_thrust}
\end{equation}
where $F_T^* = - F_X^*$ denotes the thrust force and is computed as the surface integral of pressure and shear stress over the hydrofoil.

\subsubsection{Pitch amplitude for irregular motion} \label{pitch_amp_calc}

The pitch amplitude is a key output for WAP systems since it determines the overall performance of the WAP foil. Here, the pitch amplitude is defined on a cycle-by-cycle basis using a crest-to-trough measure. After removing initial transients and segmenting the signal into complete kinematic periods of duration $T$ (section \ref{param_space_s}), the pitch angle time history within the $k^{th}$ cycle is denoted by $\theta_k(t)$. The corresponding cycle-based pitch amplitude is then defined as,
\begin{equation}
\theta_{0,k} = \frac{\max(\theta_k) - \min(\theta_k)}{2},
\end{equation}
where $\max(\theta_k)$ and $\min(\theta_k)$ are the maximum and minimum pitch angles attained within the $k^{th}$ cycle, respectively. The mean pitch amplitude, $\theta_0$ is then obtained by averaging over all retained cycles as follows,
\begin{equation}
\theta_0 = \frac{1}{N_{\mathrm{cycles}}} \sum_{k=1}^{N_{\mathrm{cycles}}} \theta_{0,k},
\label{pitch_amp_eqn}
\end{equation}
where $N_{\mathrm{cycles}}$ is the total number of complete cycles after removal of the initial transient and truncation to an integer number of periods.

\section{Results}

We present the hydrodynamic response of the hydrofoil over the parametric space summarized in Table \ref{params_table}, considering both the kinematic response of hydrofoil and the resulting flow dynamics. The effects of unimodal and bimodal wave forcing on thrust performance are also compared.

\begin{figure}[h!]
\centering
\includegraphics[angle=0, width=0.8\textwidth]{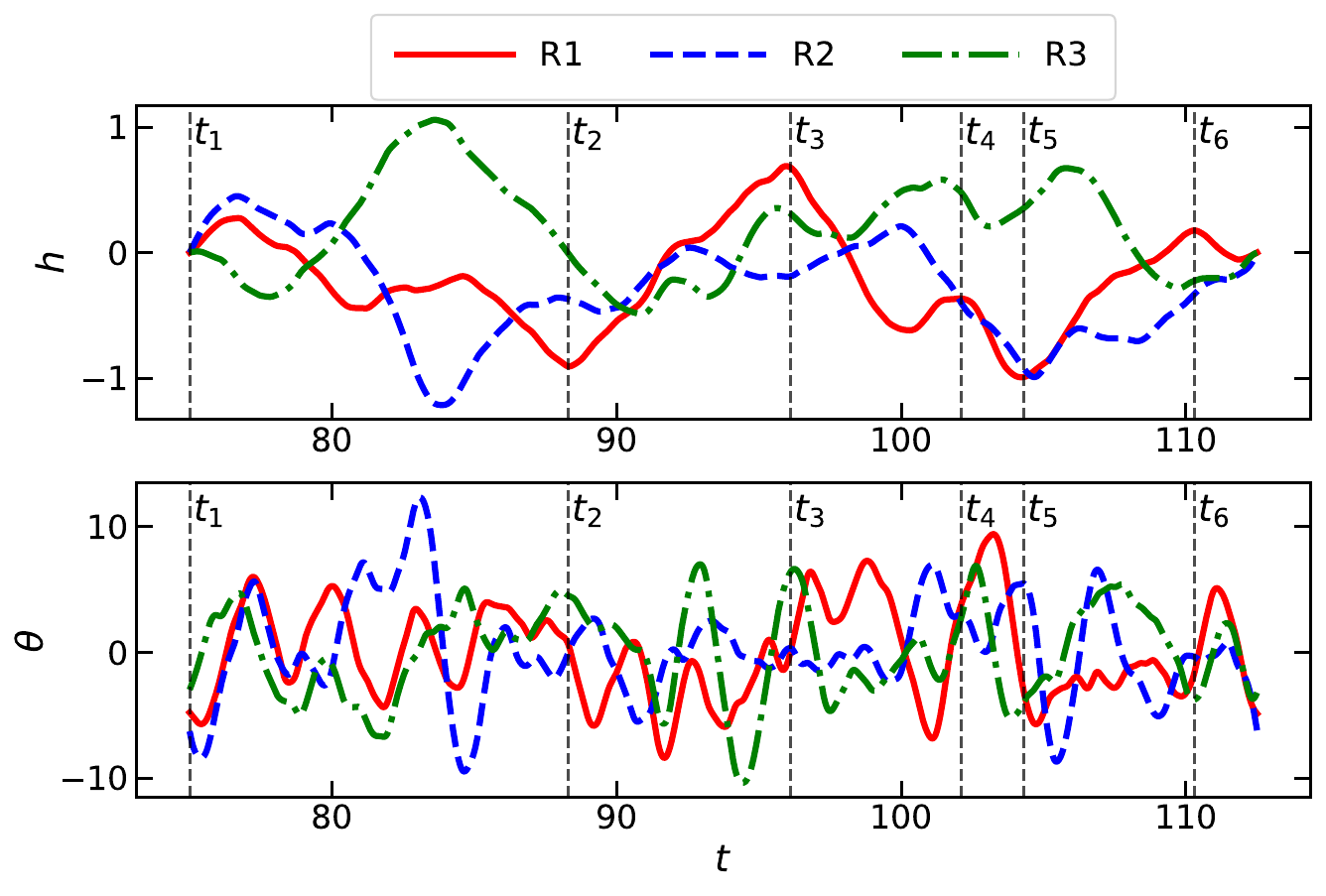}
\caption{Prescribed heave displacement, $h$, and corresponding flow-induced pitch response, $\theta$, for the representative calm sea state SS1a ($\lambda_E=9$, $f_\theta/f_{\mathrm{eff}}=6$) for three independent random phase realizations.}
\label{signals_ss1a}
\end{figure}

\subsection{Representative response of the WAP system} \label{typ_res}

Figure \ref{signals_ss1a} shows the prescribed heave motion, $h$, and the flow-induced pitch response, $\theta$, for a representative calm sea state case (SS1a, $\lambda_E=9$, $f_\theta/f_{\mathrm{eff}}=6$) across three independent random phase realizations, $R_1, R_2$ and $R_3$. While the time histories differ due to stochastic phase variations, the realizations exhibit comparable amplitude envelopes. The phase relationship between $h$ and $\theta$ varies in time as a result of the passive spring limiter dynamics. Despite being completely passive, the pitching motion remains bounded and consistent in amplitude across the three realizations. This indicates a robust performance of the passive elastic limiting mechanism under irregular bimodal forcing.
\begin{figure}[h!]
\centering
\includegraphics[angle=0, width=0.95\textwidth]{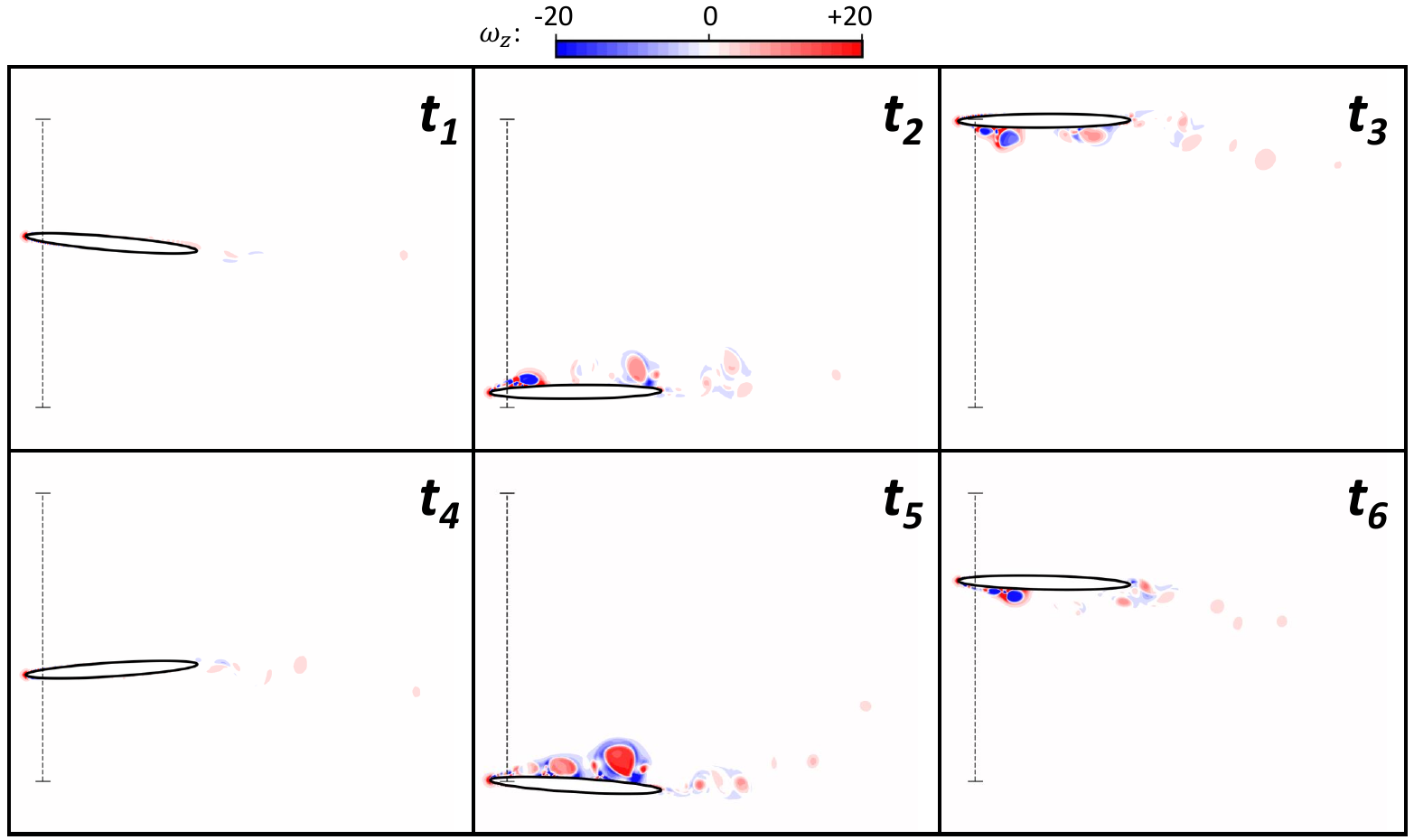}
\caption{Instantaneous wake and corresponding hydrofoil configuration at the six representative time instants ($t_1$--$t_6$) indicated in Figure \ref{signals_ss1a} for SS1a ($\lambda_E=9$, $f_\theta/f_{\mathrm{eff}}=6$). Corresponding movie is provided in the supplementary data as \texttt{movie1.mp4}.}
\label{vorticity_ss1a_rep}
\end{figure}

Furthermore, Figure \ref{vorticity_ss1a_rep} shows instantaneous vorticity fields and corresponding foil configurations at six representative time instants ($t_1$--$t_6$), as marked in Figure \ref{signals_ss1a}. Leading-edge vortices (LEVs) form alternately on both sides of foil during an oscillation cycle, and subsequently convect downstream.  As observed qualitatively through these snapshots, LEVs primarily govern the passive pitching under the influence of irregular bimodal forcing, similar to previous observations on irregular unimodal forcing \cite{raut2025hydrodynamics}.

\subsection{Propulsive performance of WAP system} \label{propulsive_performance_S}
We now evaluate thrust generation of the WAP system across three distinct sea states, as summarized in Table \ref{params_table}. For each sea state and irregular wave-based Strouhal number, $\text{St}_w$, multiple values of spring limiter frequency ratio, $f_\theta / f_{\mathrm{eff}}$, are tested. At each $f_\theta / f_{\mathrm{eff}}$, three realizations ($R_1$, $R_2$, and $R_3$) are simulated, and the cycle-averaged thrust coefficient, $\overline{C}_T$, is computed for each realization as a measure of propulsive performance.

\subsubsection{Swell-dominated sea state} \label{calm_S}

\begin{figure}[h!]
\centering
\includegraphics[angle=0, width=0.95\textwidth]{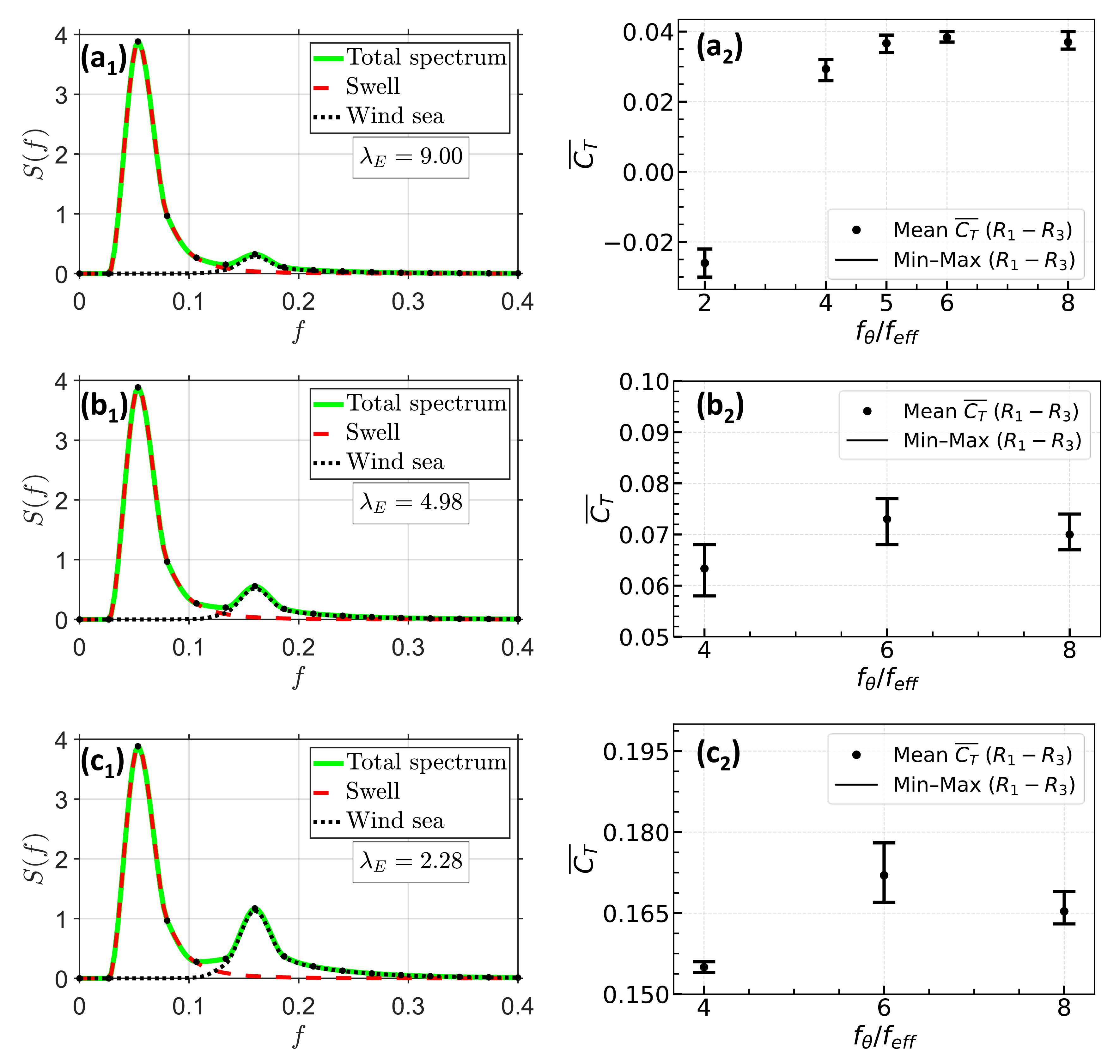}
\caption{(a$_1$--c$_1$) Coincident swell, wind-sea, and total energy spectra for the three swell-dominated sea states SS1a, SS1b, and SS1c, corresponding to $\lambda_E = 9$, $4.98$, and $2.28$, respectively. (a$_2$--c$_2$) Corresponding time-averaged thrust coefficient, $\overline{C}_T$, as a function of the spring limiter frequency ratio, $f_\theta/f_{\mathrm{eff}}$. For each $f_\theta/f_{\mathrm{eff}}$, bars indicate the range of $\overline{C}_T$ across realizations ($R_1$, $R_2$ and $R_3$), while the circular markers denote the average of $\overline{C}_T$ values over all realizations.}
\label{ss1_spectrum_Ct}
\end{figure}

Figure \ref{ss1_spectrum_Ct} shows the propulsive performance of the hydrofoil in three swell-dominated calm sea states (SS1a--SS1c). $\overline{C}_T$ is shown as a function of $f_\theta / f_\mathrm{eff}$. The reverse von Kármán wakes (Figure \ref{vorticity_ss1a_rep}) lead to positive thrust generation, as seen in Figure \ref{ss1_spectrum_Ct}($a_2-c_2$). In all cases, an optimum $f_\theta/f_{\mathrm{eff}}$ exists at which $\overline{C}_T$ attains its peak, suggesting that an appropriate elastic tuning maximizes thrust. In particular, for SS1a, SS1b, and SS1c, the maximum thrust occurs near $f_\theta/f_{\mathrm{eff}}\approx6$ in all three cases. Notably, magnitude of the maximum attainable thrust increases substantially across these cases, with the peak $\overline{C}_T$ rising from approximately $0.04$ in SS1a to $0.17$ in SS1c. This improvement follows the increase in $\text{St}_w$, suggesting that stronger wave forcing enhances propulsion performance.

\subsubsection{Comparison with unimodal forcing} \label{unimodal_vs_bimodal_s}
\begin{figure}[h!]
\centering
\includegraphics[angle=0, width=0.8\textwidth]{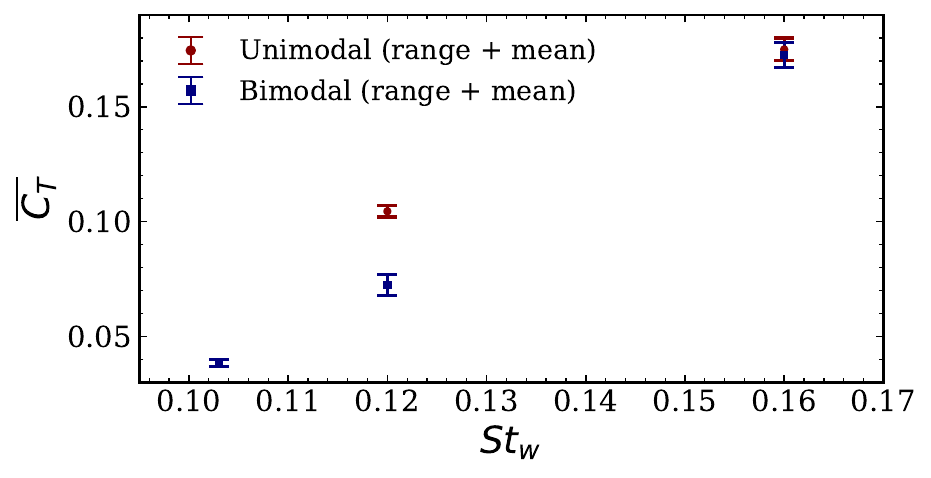}
\caption{Comparison of the maximum $\overline{C}_T$ obtained over all tested $f_\theta / f_\mathrm{eff}$ values as a function of $\text{St}_w$ for unimodal and bimodal wave forcing conditions. Bars indicate the range of $\overline{C}_T$ across realizations, while the circular markers denote the average over all realizations. Red bars represent unimodal results reported in Raut et al. \cite{raut2025hydrodynamics}, while blue bars represent bimodal results from the present work.}
\label{unimodal_vs_bimodal_f}
\end{figure}

To assess the influence of spectral bimodality, we compare the maximum thrust obtained for the three calm sea states with previously reported results for unimodal, irregular wave forcing by Raut et al. \cite{raut2025hydrodynamics}. For a direct comparison, we evaluate maximum $\overline{C}_T$ at the same values of $\text{St}_w$, as shown in Figure \ref{unimodal_vs_bimodal_f}. The bimodal cases follow the same overall increasing trend with $\text{St}_w$ and lie within the performance envelope established by the unimodal simulations. Although slight deviations are observed in $\overline{C}_T$ at very low $\text{St}_w$ values, the two datasets are close. This suggests that introducing a second spectral peak does not fundamentally degrade the propulsive capability of the WAP system. Instead, $\text{St}_w$ remains an effective defining parameter of propulsion performance even under more realistic bimodal sea conditions.

\subsubsection{Transitional and wind-sea dominated sea states} \label{transition_storm_s}
\begin{figure}[h!]
\centering
\includegraphics[angle=0, width=\textwidth]{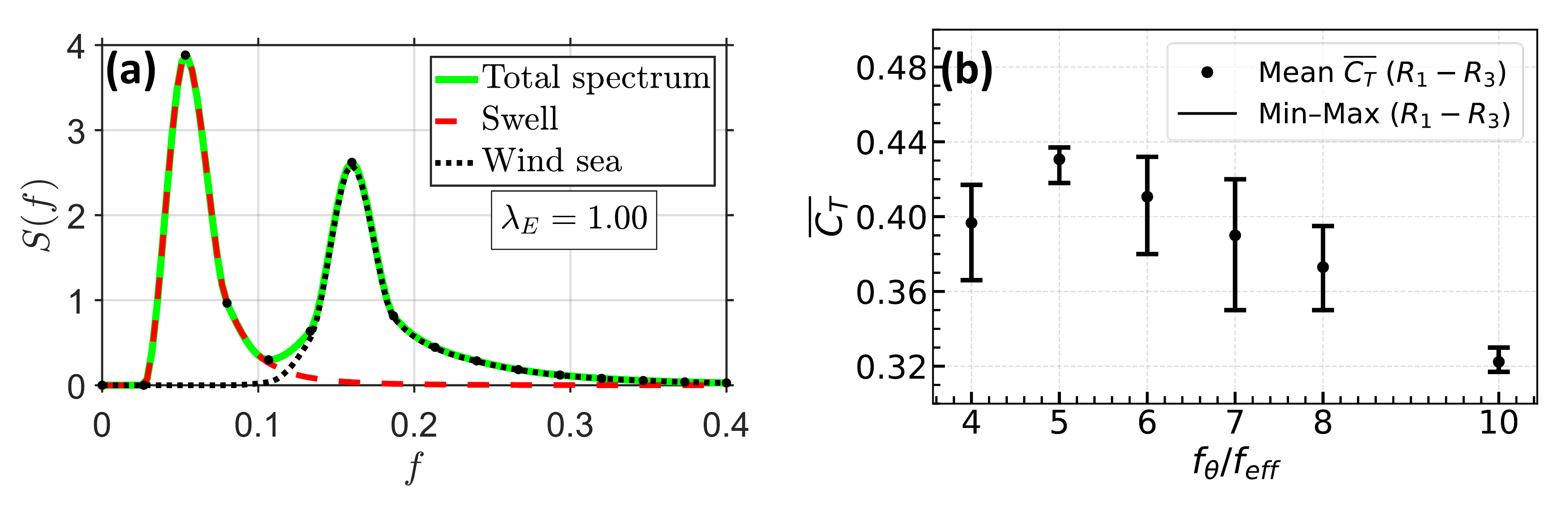}
\caption{Results for the transitional sea state with equal spectral energy of swell and wind, $\lambda_E = 1$. Rest of the caption is same as Figure \ref{ss1_spectrum_Ct}.}
\label{ss2_spectrum_Ct}
\end{figure}

\begin{figure}[h!]
\centering
\includegraphics[angle=0, width=\textwidth]{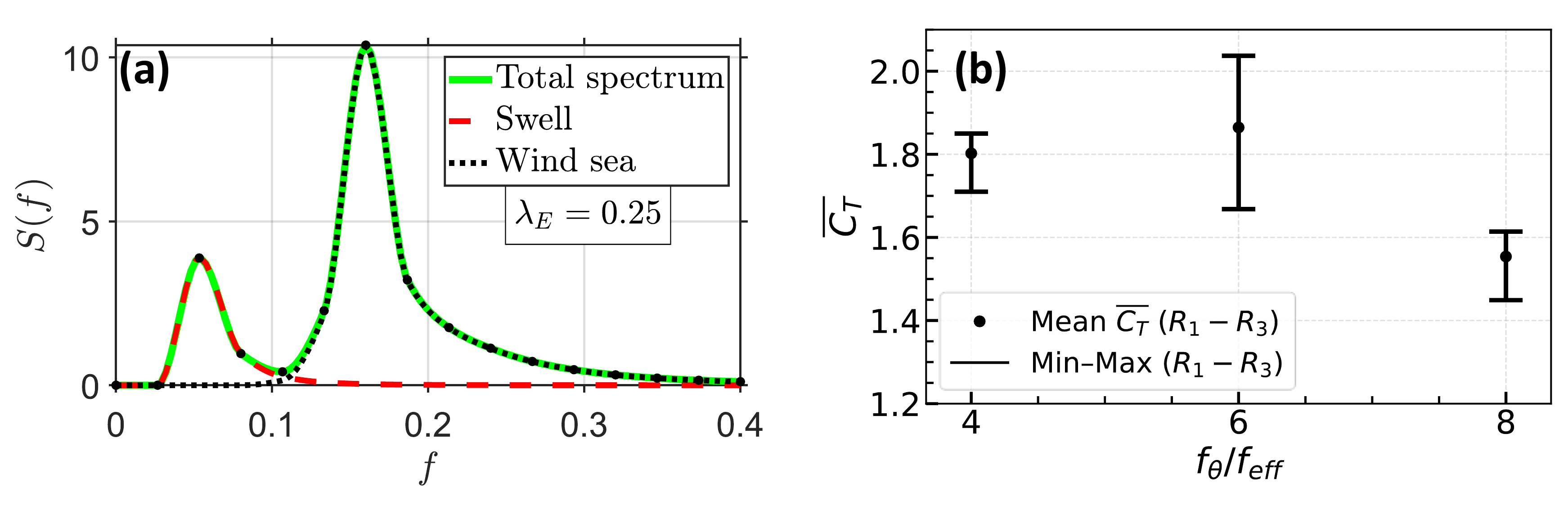}
\caption{Results for the wind-sea dominated storm sea state, $\lambda_E = 0.25$. Rest of the caption is same as Figure \ref{ss1_spectrum_Ct}.}
\label{ss3_spectrum_Ct}
\end{figure}

In the previously examined swell-dominated sea states, the influence of the high frequency wind-sea component is negligible. However, in realistic ocean environments, the wind-sea contribution can be substantial. We therefore evaluate the foil performance under two higher sea states with significant wind-sea content, spanning a broader range of ocean environments. In particular, the wind-sea spectral energy is equal to the swell energy in the transitional sea state and exceeds it in the storm sea state, as seen in Figures \ref{ss2_spectrum_Ct}(a) and \ref{ss3_spectrum_Ct}(a). 

The propulsion characteristics under the transitional (SS2) and the wind-sea dominated storm sea state (SS3) are presented in Figures \ref{ss2_spectrum_Ct} and \ref{ss3_spectrum_Ct}, respectively. Similar to the swell-dominated conditions, the hydrofoil exhibits a distinct optimum thrust with respect to $f_\theta / f_{\mathrm{eff}}$ in both higher sea states. It further confirms that passive hydroelastic tuning remains a key design parameter across a broad range of wave spectral energy distributions.

For SS2, the maximum thrust occurs near $f_\theta/f_{\mathrm{eff}} \approx 5$ ($\overline{C}_T = 0.44$, Figure \ref{ss2_spectrum_Ct}(b)), while for SS3 the optimum shifts slightly to approximately $f_\theta/f_{\mathrm{eff}} \approx 6$ ($\overline{C}_T = 1.86$, Figure \ref{ss3_spectrum_Ct}(b)). Despite substantial changes in the relative contributions of swell and wind-sea components, the optimal elastic tuning remains in a narrow range of $f_\theta/f_{\mathrm{eff}} \sim [5, 6]$, indicating a robust performance with respect to spectral variability. This suggests that a single spring limiter configuration can maintain near optimal performance across different ocean environments with varying spectral energy ratios of swell and wind-sea. The corresponding wake evolution and flapping foil dynamics under these optimal conditions are illustrated in Supplementary data as \texttt{movie2.mp4} and \texttt{movie3.mp4}.

In addition, the maximum attainable thrust increases significantly from the calm to the transitional and storm sea states, from $\overline{C}_T \approx 0.04$ at $\text{St}_w=0.103$ to $\overline{C}_T \approx 1.86$ at $\text{St}_w=0.437$. This trend indicates that increasing spectral energy, particularly from the wind-sea component, leads to enhanced thrust generation. The corresponding increase in peak thrust with $\text{St}_w$ highlights its primary role in governing performance, over and above the relative spectral composition of the sea state. The scaling of this behavior is further discussed in Section \ref{discussion}.

\section{Discussion} \label{discussion}

\subsection{\textbf{Scaling of propulsive performance across sea states}} \label{ct_summary_s}

To obtain a unified understanding of the foil's propulsive performance and identify an appropriate scaling parameter, we compare the peak values of $\overline{C}_T$ for all five wave spectra investigated in the present study. Raut et al. \cite{raut2024hydrodynamic} showed that the propulsive performance of a flapping foil is primarily governed by the LEV strength, and proposed an LEV-based model (LEVBM) to predict thrust generation under sinusoidal flapping. This framework was subsequently extended by Raut et al. \cite{raut2025hydrodynamics} to WAP systems under irregular unimodal wave forcing. Based on the equivalent spectral energy, they identified the irregular wave-based Strouhal number, $\text{St}_w$ (Equation \ref{stw_eqn}), as the governing parameter that collapses the thrust generated under irregular wave excitation. The LEVBM accurately predicted the propulsive performance for unimodal wave spectra at low $\text{St}_w$, closer to swell-dominated sea states.

\begin{figure}[h!]
\centering
\includegraphics[angle=0, width=0.8\textwidth]{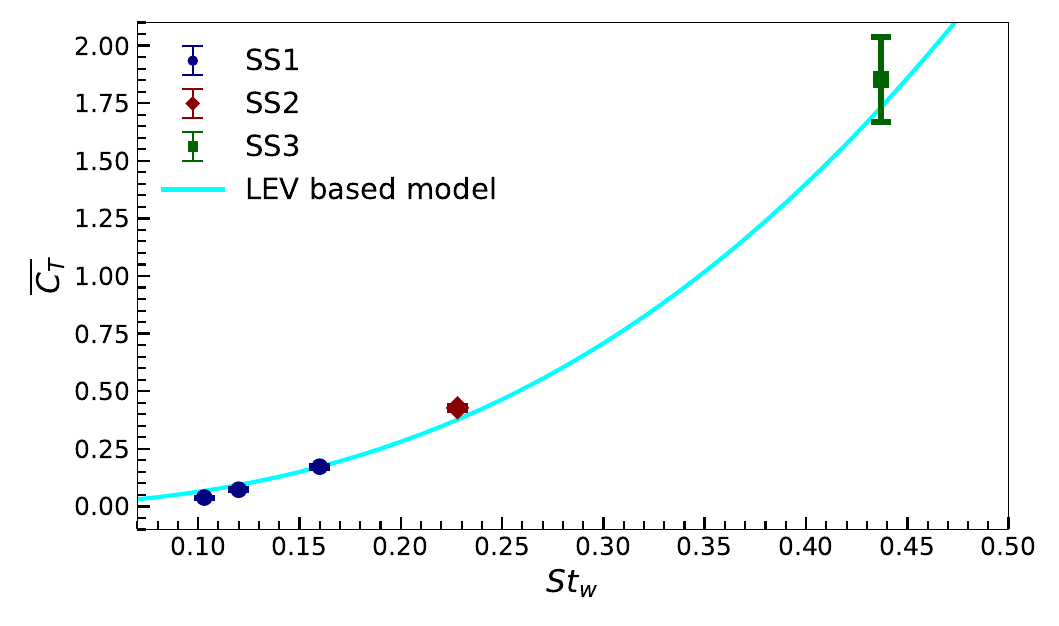}
\caption{Variation of the peak time-averaged thrust coefficient, $\overline{C}_T$ with the irregular wave-based Strouhal number, $\text{St}_w$ for all sea states tested in the present work. Markers denote the optimum thrust obtained from the present simulations, while the solid curve represents the prediction of the LEV-based model (LEVBM) proposed by Raut et al. \cite{raut2025hydrodynamics}, as described in Appendix \ref{app:LEV_model}.}
\label{stw_ct}
\end{figure}

Motivated by these findings, we now examine whether the same scaling remains valid for the more realistic bimodal wave spectra considered in the present work, which includes both swell and wind-sea components and extends to substantially higher values of $\text{St}_w$. Accordingly, Figure \ref{stw_ct} compares the maximum $\overline{C}_T$ at a given sea state (or $\text{St}_w$ value) obtained from the present simulations with the predictions of the LEVBM. We note that despite an increased spectral complexity introduced by bimodal forcing, the peak values of $\overline{C}_T$ collapse remarkably well onto the LEVBM's prediction. This agreement demonstrates that $\text{St}_w$ continues to govern the propulsive performance even when significant wind-sea energy is present.

It is worth noting that the comparison by Raut et al. \cite{raut2025hydrodynamics} was limited to unimodal wave spectra corresponding to relatively low values of $\text{St}_w$ (approximately $\text{St}_w \lesssim 0.2$), representative of swell-dominated conditions in the present work. Whether the LEV-based scaling would remain valid under ocean environments with strong wind-sea contributions was an unresolved question. The present results show that the scaling persists over the broader range of sea states considered here, including transitional and wind-sea dominated conditions. These results suggest that $\text{St}_w$ remains a primary governing parameter for WAP performance across the wide range of ocean wave environments considered here.

\subsection{\textbf{Optimal passive pitching underlies peak thrust}} \label{optimal_pitch_s}

\begin{figure}[h!]
\centering
\includegraphics[angle=0, width=0.7\textwidth]{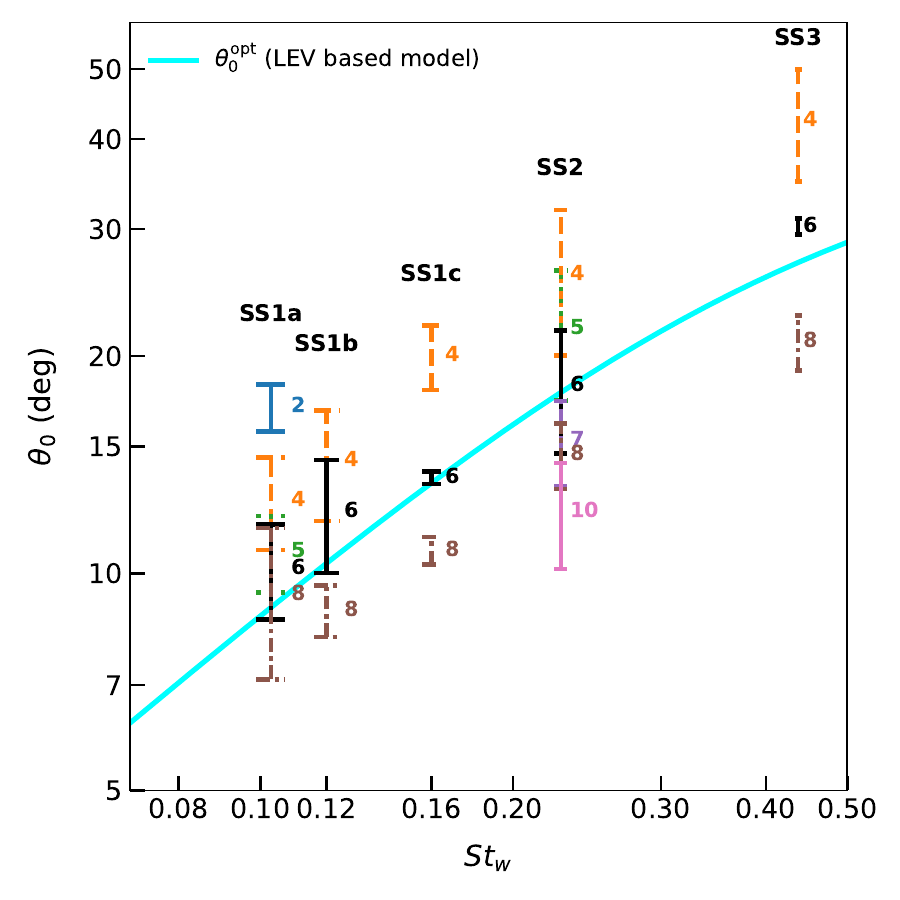}
\caption{Cycle-averaged pitch amplitude, $\theta_0$ (Equation (\ref{pitch_amp_eqn})), as a function of the irregular wave-based Strouhal number, $\text{St}_w$, for all sea states. Each column corresponds to a sea state, with its associated $\text{St}_w$ indicated on the x-axis. Within each column, the vertical bars span the minimum and maximum values of $\theta_0$ obtained from the three independent wave realizations ($R_1$, $R_2$, and $R_3$) for each tested spring limiter frequency ratio, $f_\theta/f_{\mathrm{eff}}$. Numbers adjacent to the bars identify the corresponding values of $f_\theta/f_{\mathrm{eff}}$. The solid curve shows the LEV-based model's prediction of the optimum pitch amplitude, $\theta_0^{\mathrm{opt}}$, corresponding to maximum thrust generation \cite{raut2025hydrodynamics}.}
\label{stw_pitch_angle}
\end{figure}

To elucidate the origin of peak thrust identified in Section \ref{propulsive_performance_S}, we now examine the passive pitching response of hydrofoil across all sea states. The thrust generated by a flapping foil is primarily governed by the LEV strength, which is determined by the effective angle of attack \cite{eldredge2019leading, kumar2026aerodynamic}. For a passively pitching foil, the effective angle of attack depends directly on the pitch amplitude. Based on this understanding, Raut et al. \cite{raut2025hydrodynamics} derived an LEV-based prediction for the optimum pitch amplitude, $\theta_0^{\mathrm{opt}}$, corresponding to maximum thrust generation. Accordingly, we compute the cycle-averaged pitch amplitude, $\theta_0$ (Equation (\ref{pitch_amp_eqn})), for every sea state and $f_\theta/f_{\mathrm{eff}}$, and compare it with the LEVBM's prediction of $\theta_0^{\mathrm{opt}}$ (see Appendix \ref{app:LEV_model} for details).


The results are summarized in Figure \ref{stw_pitch_angle}, which reveals a monotonic decrease in $\theta_0$ with increasing $f_\theta/f_{\mathrm{eff}}$. This trend is expected because increasing $f_\theta/f_{\mathrm{eff}}$ corresponds to increasing torsional stiffness, thereby restricting the passive pitching motion of the foil. More importantly, for all five sea states, $\theta_0$ corresponding to $f_\theta/f_{\mathrm{eff}}\approx5$--6 lie within, or very close to, the optimum range predicted by the LEV model. These are precisely the same frequency ratios at which the peak $\overline{C}_T$ was observed in Section \ref{propulsive_performance_S}.

This agreement provides a physical explanation for the peak thrust in all sea states. At low values of $f_\theta/f_{\mathrm{eff}}$, the foil is excessively compliant, producing pitch amplitudes ($\theta_0$) larger than the LEV optimum ($\theta_0^{\mathrm{opt}}$) and consequently reducing the effective angle of attack \cite{kumar2026aerodynamic}, moving it away from the condition associated with maximum LEV strength. In contrast at large values of $f_\theta/f_{\mathrm{eff}}$, the increased stiffness suppresses the pitching motion, resulting in pitch amplitudes below the optimum, leading to weaker LEV formation and therefore lower thrust. Only within a relatively narrow range of $f_\theta/f_{\mathrm{eff}}$, and therefore $\theta_0$, does the foil generate the LEV of optimal strength, which maximizes thrust. The persistence of this behavior across all sea states (swell-dominated, transitional, and wind-sea dominated) indicates that the optimum propulsion mechanism is primarily governed by the passive pitch kinematics, and is less dependent on the relative distribution of spectral wave energy.

\subsection{\textbf{WAP design guideline}} \label{design_guideline}

The preceding results provide a simple and physically meaningful guideline for the design of WAP systems under realistic ocean environments: rather than tuning the torsional stiffness for a specific sea state or dominant wave frequency, the objective should be to select the hydroelastic properties such that the passive pitching response remains close to the LEV-optimal pitch amplitude.

The results presented in Sections \ref{propulsive_performance_S}, \ref{ct_summary_s} and \ref{optimal_pitch_s} collectively demonstrate that this objective can be achieved by tuning the spring limiter's torsional natural frequency to approximately $5 \lesssim {f_\theta}/{f_{\mathrm{eff}}} \lesssim 6$, where $f_{\mathrm{eff}}$ is the effective peak spectral frequency computed as the weighted sum of swell and wind-sea frequencies (Equation \ref{eff_freq_eq}). Within this range, the foil attains the LEV-optimal passive pitch amplitude across a wide range of sea states, resulting in maximum thrust generation despite substantial variation in spectral contributions of swell and wind-sea waves. These observations further indicate that the effective peak spectral frequency, $f_\mathrm{eff}$, provides a single characteristic time scale that accounts for the combined effects of swell and wind-sea components. Consequently, the hydroelastic tuning of a passive WAP system can be based primarily on the effective peak spectral frequency, eliminating the need to treat the swell and wind-sea spectral peaks separately. This considerably simplifies the design procedure of WAP systems for realistic ocean environments.

\section{Conclusions}\label{conclusion} 
We have examined the performance of a submerged hydrofoil wave-assisted propulsion (WAP) system operating in realistic bimodal sea states generated by the coexistence of swell and wind-sea components. High-fidelity fluid--structure interaction simulations spanning calm, transitional, and storm conditions demonstrate that WAP systems remain capable of generating robust propulsive thrust over the wide range of bimodal sea states considered here. Despite the increased complexity of bimodal wave forcing, the propulsion characteristics were found to follow the same effective peak spectral frequency scaling previously established for monochromatic and unimodal seas, confirming the broader applicability of these scaling laws and providing a unified framework for characterizing WAP performance across diverse sea states.

The simulations also demonstrate that hydroelastic tuning plays a central role in maximizing thrust generation. 
The optimal torsional spring stiffness varies with sea-state characteristics, primarily through changes in the effective peak spectral frequency, while the optimal normalized frequency ratio remains approximately $f_\theta/f_{\mathrm{eff}} \sim [5,6]$. Thus, a fixed torsional stiffness may not remain optimally tuned as the sea state characteristics change. These observations therefore motivate the development of controllable or adaptive torsional spring mechanisms capable of tuning the hydrofoil response to changing sea conditions.

Overall, the present study extends the understanding of wave-assisted propulsion from idealized wave environments to the more complex bimodal sea states frequently encountered in the open ocean. The results demonstrate both the robustness of WAP systems and the importance of adaptive hydroelastic tuning, providing practical design guidelines for next-generation wave-powered marine propulsion technologies.

\section{CRediT authorship contribution statement}
\textbf{Avinash Kumar Pandey}: Writing – review \& editing, Writing – original draft, Visualization, Validation, Methodology, Investigation, Conceptualization. \textbf{Jung-Hee Seo}: Writing – review \& editing, Supervision, Resources, Investigation, Funding acquisition, Conceptualization. \textbf{Rajat Mittal}: Writing – review \& editing, Supervision, Resources, Project administration, Funding acquisition, Conceptualization.

\section{Declaration of competing interest}
The authors declare that they have no known competing financial interests or personal relationships that could have appeared to
influence the work reported in this paper

\section{Acknowledgments}
We gratefully acknowledge the High Performance Computing (HPC) facility \emph{Rockfish} at Johns Hopkins University for providing the computational resources used to carry out the simulations in this work. We also thank H. S. Raut for helpful discussions.

\section{Data availability}
The data that support the findings of this study are available from the authors upon reasonable request.

\section{Supplementary data}\label{supp_data}

Illustrative supplementary movies showcasing the transient wake dynamics and corresponding flapping foil motions for representative cases are provided below:

\noindent \textbf{\textit{movie1.mp4:}} Transient wake evolution and corresponding hydrofoil motion under a calm sea state (SS1a, $\text{St}_w=0.103$), with $\lambda_E=9$ and $f_\theta/f_{\mathrm{eff}}=6$. The vorticity field is visualized using a colormap range of $\omega_z\in[-20,20]$.

\noindent \textbf{\textit{movie2.mp4:}} Transient wake evolution and corresponding hydrofoil motion under a transitional sea state (SS2, $\text{St}_w=0.228$), with $\lambda_E=1$ and $f_\theta/f_{\mathrm{eff}}=5$. The vorticity field is visualized using a colormap range of $\omega_z\in[-20,20]$.

\noindent \textbf{\textit{movie3.mp4:}} Transient wake evolution and corresponding hydrofoil motion under a storm sea state (SS3, $\text{St}_w=0.437$), with $\lambda_E=0.25$ and $f_\theta/f_{\mathrm{eff}}=6$. The vorticity field is visualized using a colormap range of $\omega_z\in[-50,50]$.

\section{Appendix}
\subsection{Leading-edge vortex (LEV)-based model (LEVBM)}
\label{app:LEV_model}

The leading-edge vortex-based model (LEVBM) developed by Raut et al. \cite{raut2025hydrodynamics}
provides a reduced-order description of thrust generation by relating the LEV-induced force to the kinematics of the flapping foil. The model assumes that the LEV strength ($\Lambda_{\mathrm{LEV}}$) is proportional to the effective angle of attack ($\alpha_{\mathrm{eff}}$) as follows,
\begin{equation}
\Lambda_{\mathrm{LEV}}(t)
=
\sin(\alpha_{\mathrm{eff}}(t))\sin(\theta(t)).
\label{eq:LEV_parameter}
\end{equation}

The thrust coefficient is assumed to scale linearly with $\Lambda_{\mathrm{LEV}}$. Consequently, the optimum pitch amplitude is obtained by maximizing $\Lambda_{\mathrm{LEV}}$ as follows,
\begin{equation}
\theta_{0,\mathrm{opt}}
=
\underset{\theta_0}{\operatorname{arg\,max}}
\left(
\overline{\Lambda}_{\mathrm{LEV}}
\right).
\label{eq:theta_opt_general}
\end{equation}

For harmonic wave-induced heaving motion, this optimization yields the following analytical
relationship,

\begin{equation}
\theta_{0,\mathrm{opt}}
=
\frac{1}{2}\tan^{-1}(\pi \text{St}_w)-\theta_s ,
\label{eq:theta_opt}
\end{equation}
where $\theta_s$ is a small correction associated with the foil leading-edge geometry and
can be neglected for a thin elliptic foil. The wave-based Strouhal number is defined as,

\begin{equation}
St_w=\frac{2fh_0}{U_\infty},
\label{eq:St_regular}
\end{equation}
where $f$ and $h_0$ are the wave frequency and heave amplitude, respectively. The optimal thrust is then obtained by substituting
$\theta_{0,\mathrm{opt}}$ into the LEVBM equations and evaluating the cycle-averaged
thrust coefficient as follows,

\begin{equation}
\overline{C}_{T,\mathrm{opt}}
=
\frac{1}{T}
\int_0^T
K\Lambda_{\mathrm{LEV}}
\left(
\theta_{0,\mathrm{opt}}
\right)
dt .
\label{eq:CT_opt}
\end{equation}
where $K$ denotes a proportionality constant. For the thin elliptic hydrofoil considered here, we use the value $K=5$ adopted by Raut et al. \cite{raut2025hydrodynamics}.

For irregular waves, a unique characteristic wave frequency and amplitude cannot be defined. Therefore, the Strouhal number is instead defined using the measured RMS heave velocity of the foil, with $\dot{h}_\mathrm{rms} = \sqrt{2}\pi f h_0$ substituted into Equation (\ref{eq:St_regular}), which gives,

\begin{equation}
St_{w,\mathrm{irr}}
=
\frac{\sqrt{2}\dot{h}_{\mathrm{rms}}}
{\pi U_\infty},
\label{eq:St_irregular}
\end{equation}
where $\dot{h}_{\mathrm{rms}}$ is the ensemble-averaged root-mean-square value of the heave velocity. This definition represents the effective velocity scale of the heaving motion and allows the LEVBM scaling to be applied to non-harmonic motions.\\
\\

\newpage

\bibliographystyle{ieeetr}
\bibliography{WAP_bimodal} 

\pagebreak


\end{document}